\documentclass[twocolumn,english]{revtex4-2}
\usepackage{lmodern}
\usepackage{lmodern}

\usepackage[T1]{fontenc}
\usepackage[latin9]{inputenc}
\usepackage{babel}
\usepackage{amsmath}
\usepackage{amssymb}
\usepackage{graphicx}
\usepackage[unicode=true,pdfusetitle,
 bookmarks=true,bookmarksnumbered=false,bookmarksopen=false,
 breaklinks=false,pdfborder={0 0 1},backref=false,colorlinks=false]
 {hyperref}

\makeatletter

\providecommand{\tabularnewline}{\\}

\usepackage{babel}

\makeatother

\begin{document}
\title{Solutions of the converging and diverging shock problem in a medium
with varying density}
\author{Itamar Giron}
\author{Shmuel Balberg}
\author{Menahem Krief}
\email{menahem.krief@mail.huji.ac.il}

\affiliation{Racah Institute of Physics, The Hebrew University, 9190401 Jerusalem,
Israel}
\begin{abstract}
We consider the solutions of the Guderley problem, consisting of a
converging and diverging hydrodynamic shock wave in an ideal gas with
a power law initial density profile. The self-similar solutions, and
specifically the reflected shock coefficient, which determines the
path of the reflected shock, are studied in detail, for cylindrical
and spherical symmetries and for a wide range of values of the adiabatic
index and the spatial density exponent. Finally, we perform a comprehensive
comparison between the analytic solutions and Lagrangian hydrodynamic
simulations, by setting proper initial and boundary conditions. A
very good agreement between the analytical solutions and the numerical
simulations is obtained. This demonstrates the usefulness of the analytic
solutions as a code verification test problem. 
\end{abstract}
\maketitle

\section{Introduction \label{sec:introduction}}

The Guderley problem is a well known hydrodynamic problem of a strong
shock propagating radially in an ideal gas medium. The shock originates
at infinity and collapses to a central point or an axis, where it
is reflected and diverges towards infinity \cite{guderley1942starke,lazarus1977similarity,lazarus1981self,ramsey2012guderley,sakurai1960problem,sharma1995similarity,toque2001self,ruby2019boundary,biasi2021self}.
The problem was first formulated with an initially uniform density
profile, and was analyzed comprehensively and solved for this case
by Lazarus \textbf{and Richtmyer} \cite{lazarus1977similarity,lazarus1981self}.
The theory of converging and diverging compressible shocks is an active
research area, where for example, the Guderley problem was generalized
to include: non-ideal equations of state \cite{ramsey2018converging,singh2020kinematics,sharma2022similarity,calvo2022stability,singh2022convergence,chefranov2022limitation},
magneto-hydrodynamics \cite{chauhan2021piston,yadav2021propagation},
and gravitation \cite{nath2022propagation}. Recently, a general solution
for continuous converging waves was developed \cite{jenssen2023radially}.

Imploding shocks are often invoked as a possible ignition mechanism
in reactive material. These include chemical compounds (see Ref.  \cite{OG2007}
for a review on laboratory experiments), and thermonuclear, in inertial
confinement fusion experiments \cite{vallet2013finite,abu2022lawson}
and detonation in white dwarfs in astrophysics \cite{kushnir2012imploding,ShenBildsten2014}.
A fundamental feature in such ignition scenarios is that both the
converging and diverging shocks may lead to ignition, since the particulars
depend on a competition between the hydrodynamical time and the heating
time due to the exothermal burning processes. Interestingly, the critical
strength of the shock required to reach ignition is usually determined
by the diverging shock. While this shock is typically not strong (as
opposed to the earlier converging shock), the post-shock material
in this stage has higher densities and temperatures, having being
shocked twice; see the analytical and numerical analysis by \cite{kushnir2012imploding}.
Hence, an accurate calculation of the entire converging-diverging
shock sequence is crucial to understanding (and designing) an ignition
candidate system.

In Ref. \cite{giron2021solutions} we presented generalized solutions
for the converging shock for a power law initial density profile $\rho\left(r\right)\sim r^{\mu}$,
with a general spatial exponent $\mu$ (see also \cite{CHERNOUSKO19601334,sharma1995similarity,toque2001self,modelevsky2021revisiting}).
In this work we extend this generalization of the initial density
profile, to the reflected diverging shock solution. By combining the
solutions for the converging and diverging flows we obtain a solution
of the complete Guderley problem for an initial power law density
profile.

The Guderley problem is a self-similar problem of the second kind,
in which the similarity is found by requiring the solution to pass
through a singularity, such as a sonic point. As a result, its solution
is given in terms of similarity profiles as the hydrodynamic Euler
equations are transformed into a system of ordinary differential equations
(ODE). An extension of the problem to a general power law initial
density is possible, since this generalization does not introduce
additional dimensional parameters to the problem \cite{zel2002physics,kamm2000evaluation,krief2021analytic,krief2023piston}.
The equations describing the similarity profiles contain two unknown
parameters which need be determined, either approximately \cite{chisnell1998analytic,vishwakarma2005analytic}
using simple analytical methods, or exactly using more sophisticated
numerical methods. An exact numerical method for the calculation of
the first parameter, the similarity exponent $\lambda$ was developed
for general $\mu$, in our previous work \cite{giron2021solutions}.
The second parameter, which is the diverging shock constant $B$,
was calculated in the literature for $\mu=0$ by Lazarus et. al \cite{lazarus1977similarity,lazarus1981self,ramsey2012guderley}
and Ramsey et. al \cite{ramsey2012guderley}. 

The Guderley problem can be used as a nontrivial test problem for
hydrodynamic simulation codes \cite{ramsey2012guderley,ramsey2012simulation,ramsey2012surrogate,ramsey2017verification,ramsey2018converging,ramsey2019piston,ruby2019boundary,giron2021solutions}.
Using the generalized semi-analytic solutions, we define two types
of test problems for hydrodynamic simulation codes, by properly defining
the initial and boundary conditions. The tests represent the dynamics
of the diverging shock as well as the dynamics of both the converging
and diverging shock in the same simulation. We perform a detailed
comparison between the simulations and the analytic solutions, for
a variety of $\gamma$ (the ideal gas adiabatic index) and $\mu$.

In this work we focus on the calculation of the diverging shock. We
generalize the method of Lazarus \cite{lazarus1981self} and present
a robust method for the calculation of the diverging shock constant
$B$, for a general value of the initial density exponent $\mu$.
The structure of the article is as follows. In section \ref{sec:The-Guderley-Problem}
we review the generalized Guderley problem and its self-similar solutions
for the converging and diverging shocks are presented. In section
\ref{sec:Determining-the-Value} a robust method for the calculation
of the diverging shock constant and the resulting similarity profiles,
for general values of $\mu$ is developed. The solutions are calculated
for a wide range of values of $\gamma$ and $\mu$, and are compared
to the results of previously published works when such exist. In section
\ref{sec:Comparison-to-a} we define two types of test problems for
hydrodynamic simulation codes, by properly defining the initial and
boundary condition of a piston with velocity given from the analytic
solution \cite{ramsey2017verification}. We perform detailed hydrodynamic
simulations and compare the results to the similarity solutions. We
conclude in section \ref{sec:Summary}.

\section{The Guderley Problem \label{sec:The-Guderley-Problem}}

\begin{figure*}[t]
\begin{centering}
\includegraphics[scale=0.3]{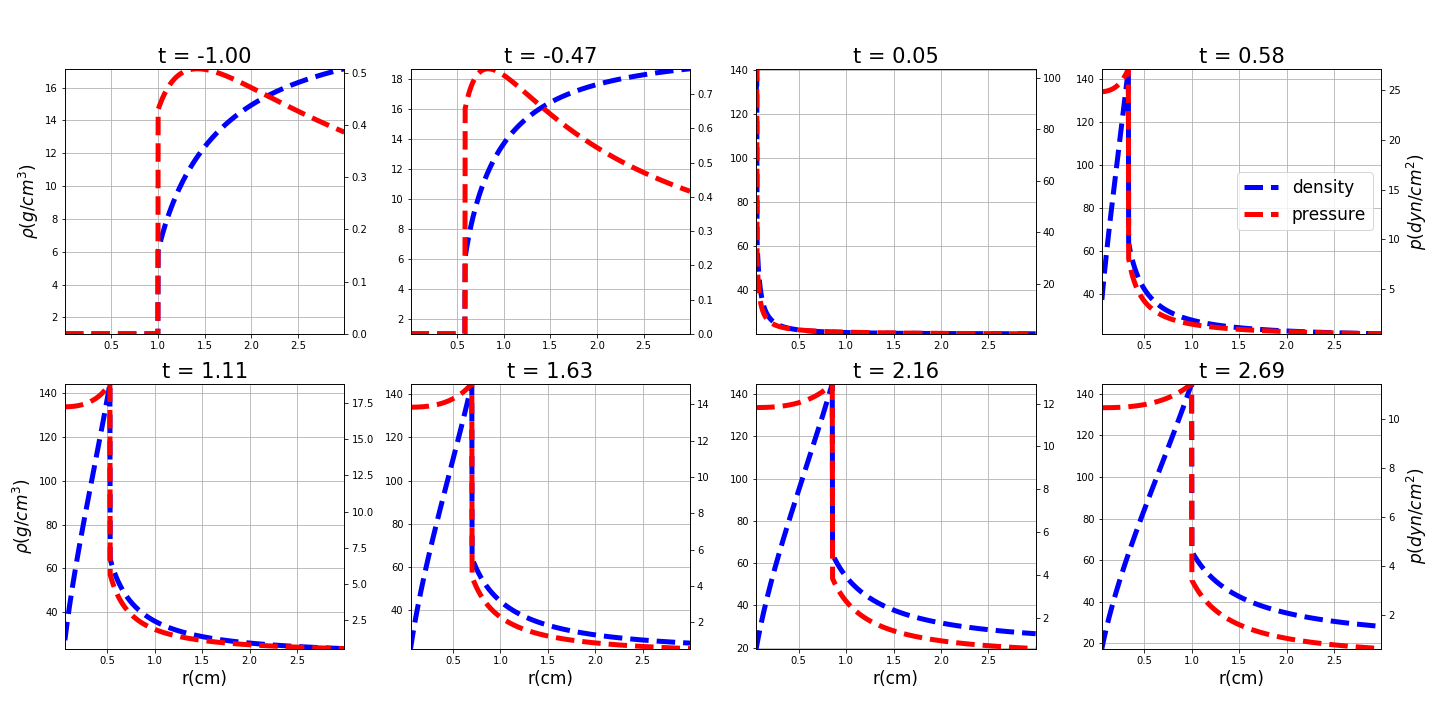}
\par\end{centering}
\caption{The density (blue) and pressure (red) profiles of the converging and
diverging shock problem, at various times which are listed in the
titles (time is ordered by: left to right, top to bottom). The converging
shock is originated at $r=\infty,\ t=-\infty$, and reaches the point
$r=1$ at time $t=-1$. Subsequently it converges at the center $r=0$
at time $t=0$ and then reflects as a diverging shock which propagates
outwards, reaching the point $r=1$ at $t=B$ (see equation \ref{eq: shock position}).
The case shown is for $\rho_{0}=1\text{g/cm}^{3}$ and the parameters
$\mu=0$ (an initially homogeneous density), $\gamma=1.4$, $n=3$
(spherical symmetry), for which $\lambda=1.3944$ and $B=2.6885$.
\label{fig:The-analytical-density-pressure}}
\end{figure*}

\begin{figure}
\begin{centering}
\includegraphics[scale=0.6]{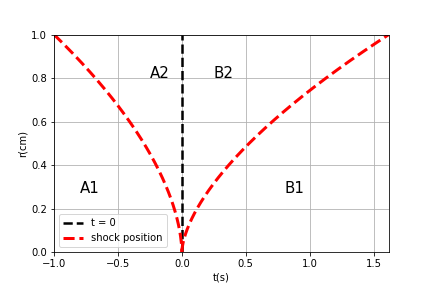}
\par\end{centering}
\caption{The converging and diverging shock positions as a function of time,
for the case $\gamma=2,$ $\mu=1,$ $n=3$, for which $\lambda=1.7498,\ B=1.6189$.
The $r-t$ plane is split into four regions: (\textbf{A1})\textbf{
}is the undisturbed region in the converging shock case, (\textbf{A2})
is the shocked region in the converging shock problem, (\textbf{B1})\textbf{
}is the region shocked by the diverging shock and (\textbf{B2})\textbf{
}is the region in front of the diverging shock. In the previous work
\cite{giron2021solutions} the incoming shock regions \textbf{A1,A2}
were examined. This work focuses on the outgoing shock regions \textbf{B1,B2}.\label{fig:analytic_shock_pos}}
\end{figure}

In this work we consider the hydrodynamic problem of a shock wave
which diverges to infinity in an ideal gas medium. This diverging
shock is originated by a previously converging shock wave according
to the Guderley problem, which has reached the origin and reflected
from a central point (spherical symmetry) or an axis (cylindrical
symmetry), as demonstrated in figures \ref{fig:The-analytical-density-pressure}-\ref{fig:analytic_shock_pos}.

For completeness, we list again the governing equations and employ
the same notation as in Ref. \cite{giron2021solutions}. The Euler
equations which describe the compressible gas flow are:

\begin{equation}
\frac{1}{\rho}\frac{\partial\rho}{\partial t}+\frac{u}{r}\frac{\partial\rho}{\partial r}+\frac{\partial u}{\partial r}+\left(n-1\right)\frac{u}{r}=0,\label{eq:continuity}
\end{equation}
\begin{equation}
\frac{\partial u}{\partial t}+u\frac{\partial u}{\partial r}+\frac{c^{2}}{\gamma\rho}\frac{\partial\rho}{\partial r}+2c\frac{\partial c}{\partial r}=0,
\end{equation}
\begin{equation}
\frac{\partial c}{\partial t}+u\frac{\partial c}{\partial r}+\frac{\left(\gamma-1\right)c}{2}\left(\frac{\partial u}{\partial r}+\frac{\left(n-1\right)u}{r}\right)=0,\label{eq:energy_cons}
\end{equation}
where $\rho\left(r,t\right)$ denotes the fluid density, $u\left(r,t\right)$
the velocity, $c\left(r,t\right)$ the sound speed, and $n$ the symmetry
dimension ($n=2$ for cylindrical axial symmetry and $n=3$ for spherical
symmetry). The medium is assumed to consist of an ideal gas with an
adiabatic index $\gamma$, so that the equation of state is: 
\begin{equation}
p\left(\rho,e\right)=\left(\gamma-1\right)\rho e,
\end{equation}
relating the pressure $p$ to the specific internal energy $e$ and
density $\rho$. The sound speed is then: 
\begin{equation}
c=\sqrt{\frac{\gamma p}{\rho}}.
\end{equation}
A converging Guderley shock is originated at $r=\infty$ and $t=-\infty$,
and propagates in a cold material which is at rest and whose initial
density profile has a spatial power law of the form \cite{giron2021solutions}:
\begin{equation}
\rho\left(r,t=-\infty\right)=\rho_{0}r^{\mu}.\label{eq:rhopowlaw}
\end{equation}
We note that for $\mu<0$ the density diverges at $r=0$, and henceforth
we limit ourselves to the constraint $\mu\geq-n$ in order for the
total mass enclosed by finite radii to be finite as well. It is noteworthy
that this choice limits us to the physical regime where the converging
shock must reach the center; there does exist a critical value $\mu_{b}<-n$
for which the shock does not converge due to the step density gradient
towards the origin. Similarly, we also limit our analysis to values
$\mu<\mu_{c}$, which is a positive critical value, for which steeper
outward-increasing density profiles will cause the diverging shock
to stall. See \cite{modelevsky2021revisiting} for a derivation of
$\mu_{b},\mu_{c}$ as a function of dimension and $\gamma$.

The Rankine--Hugoniot jump relations relate the values of the thermodynamic
quantities across the shock: 
\begin{equation}
D\left(\rho_{2}-\rho_{1}\right)=\rho_{2}u_{2}-\rho_{1}u_{1},\label{eq:massjump}
\end{equation}
\begin{equation}
p_{1}+\rho_{1}\left(u_{1}-D\right)^{2}=p_{2}+\rho_{2}\left(u_{2}-D\right)^{2},\label{eq:momentumjump}
\end{equation}
\begin{equation}
e_{1}+\frac{p_{1}}{\rho_{1}}+\frac{1}{2}\left(u_{1}-D\right)^{2}=e_{2}+\frac{p_{2}}{\rho_{2}}+\frac{1}{2}\left(u_{2}-D\right)^{2},\label{eq:energyjump}
\end{equation}
where $D$ is the shock velocity, and the indices 1 and 2 correspond
to the pre-shocked and post-shocked components, respectively. The
solution of the converging shock is described via a self-similar representation
at negative times, so that incoming shock position is given as a temporal
power law:

\begin{equation}
r_{\text{in-shock}}\left(t\right)=\left(-t\right)^{\frac{1}{\lambda}},\label{eq:in_shock_position}
\end{equation}
where $\lambda$ is the similarity exponent. The independent similarity
variable $x$, is defined by:

\begin{equation}
x=\frac{t}{r^{\lambda}}.\label{eq:xdef}
\end{equation}
The similarity functions for the velocity, sound-speed and density
profiles, $V\left(x\right)$, $C\left(x\right)$ and $R\left(x\right)$
are defined, respectively, as:

\begin{equation}
u\left(r,t\right)=-\frac{r}{\lambda t}V\left(x\right),\label{eq:V_def}
\end{equation}
\begin{equation}
c\left(r,t\right)=-\frac{r}{\lambda t}C\left(x\right),\label{eq:C_def}
\end{equation}
\begin{equation}
\rho\left(r,t\right)=\rho_{0}r^{\mu}R\left(x\right).\label{eq:R_def}
\end{equation}
Substituting of the self-similar representation \eqref{eq:V_def}-\eqref{eq:R_def}
in the Euler equations \eqref{eq:continuity}-\eqref{eq:energy_cons},
results in a system of ordinary differential equations (ODEs) for
the similarity profiles :

\begin{equation}
\boldsymbol{A}\frac{d}{dx}\left[\begin{array}{c}
R\\
V\\
C
\end{array}\right]=\left[\begin{array}{c}
V(\mu+n)\\
R\left(C^{2}(2+\mu)+\gamma V(\lambda+V)\right)\\
\frac{C\left(2\left(V+\lambda\right)+n(\gamma-1)\right)}{2\gamma(1+V)}
\end{array}\right],\label{eq:linsys}
\end{equation}
where $\boldsymbol{A}$ is the matrix: 
\[
\boldsymbol{A}=\lambda x\left[\begin{array}{ccc}
\frac{1+V}{R} & 1 & 0\\
C^{2} & \gamma R(V+1) & 2RC\\
0 & \frac{C(\gamma-1)}{2\gamma(V+1)} & \frac{1}{\gamma}
\end{array}\right].
\]
Finally, solving for the derivatives gives the canonical ODE form:
\begin{equation}
\lambda xR'=\frac{\Delta_{1}}{\Delta},\ \ \lambda xV'=\frac{\Delta_{2}}{\Delta},\ \ \lambda xC'=\frac{\Delta_{3}}{\Delta},\label{eq:ODEs}
\end{equation}
with the following determinants:

\begin{equation}
\Delta\left(V,C\right)=C^{2}-\left(V+1\right)^{2},\label{eq:delta}
\end{equation}
\begin{align}
\Delta_{1}(R,V,C,\lambda,\mu) & =R\Bigg[\frac{2(1-\lambda)+\mu(\gamma V+1)}{\gamma(V+1)}C^{2}\nonumber \\
 & +V(V+\lambda)-(n+\mu)V(V+1)\Bigg],
\end{align}

\begin{align}
\Delta_{2}(V,C,\lambda,\mu) & =C^{2}\left(nV+\frac{2(\lambda-1)-\mu}{\gamma}\right)\nonumber \\
 & \ \ \ \ -V(V+1)(V+\lambda),\label{eq:delta2}
\end{align}
\begin{align}
\Delta_{3}(V,C,\lambda,\mu) & =C\Bigg[C^{2}\left(1+\frac{2(\lambda-1)+\mu(\gamma-1)}{2\gamma(1+V)}\right)\nonumber \\
 & \ \ \ \ -(V+1)^{2}-(n-1)(\gamma-1)\frac{V(1+V)}{2}\nonumber \\
 & \ \ \ \ -(\lambda-1)\frac{(3-\gamma)V+2}{2}\Bigg].\label{eq:delta3}
\end{align}
We note that the typographical error in equation 16 of Ref. \cite{giron2021solutions}
is now corrected in equation \ref{eq:delta} (the order of the terms
on the right hand side was corrected).

The imploding shock reaches the center at time $t=0$, and is then
reflected into a diverging shock which propagates outward. As shown
by Lazarus in Refs. \cite{lazarus1981self,lazarus1977similarity},
the diverging shock position has the same similarity exponent $\lambda$,
so that the converging and diverging shock positions are written as
(see figure \ref{fig:analytic_shock_pos}):

\begin{equation}
r_{\text{shock}}\left(t\right)=\begin{cases}
\left(-t\right)^{\frac{1}{\lambda}} & t\leq0\\
\left(\frac{t}{B}\right)^{\frac{1}{\lambda}} & t>0
\end{cases}\label{eq: shock position}
\end{equation}
where $B=B\left(\gamma,\mu,n\right)$ is a constant which represents
the time it takes the outgoing shock to travel from $r=0$ to $r=1$,
and depends only on the defining quantities in the problem: $\gamma,\mu,n$.
In Ref. \cite{giron2021solutions} we discussed in detail how to calculate
the similarity exponent $\lambda\left(\gamma,\mu,n\right)$ using
iterative methods. In the following we complete the derivation with
an analytic solution for the diverging shock, for which we develop
in detail a method for calculating $B\left(\gamma,\mu,n\right)$.

\section{Calculation of $B\left(\gamma,\mu,n\right)$ \label{sec:Determining-the-Value}}

\begin{figure}
\begin{centering}
\includegraphics[scale=0.45]{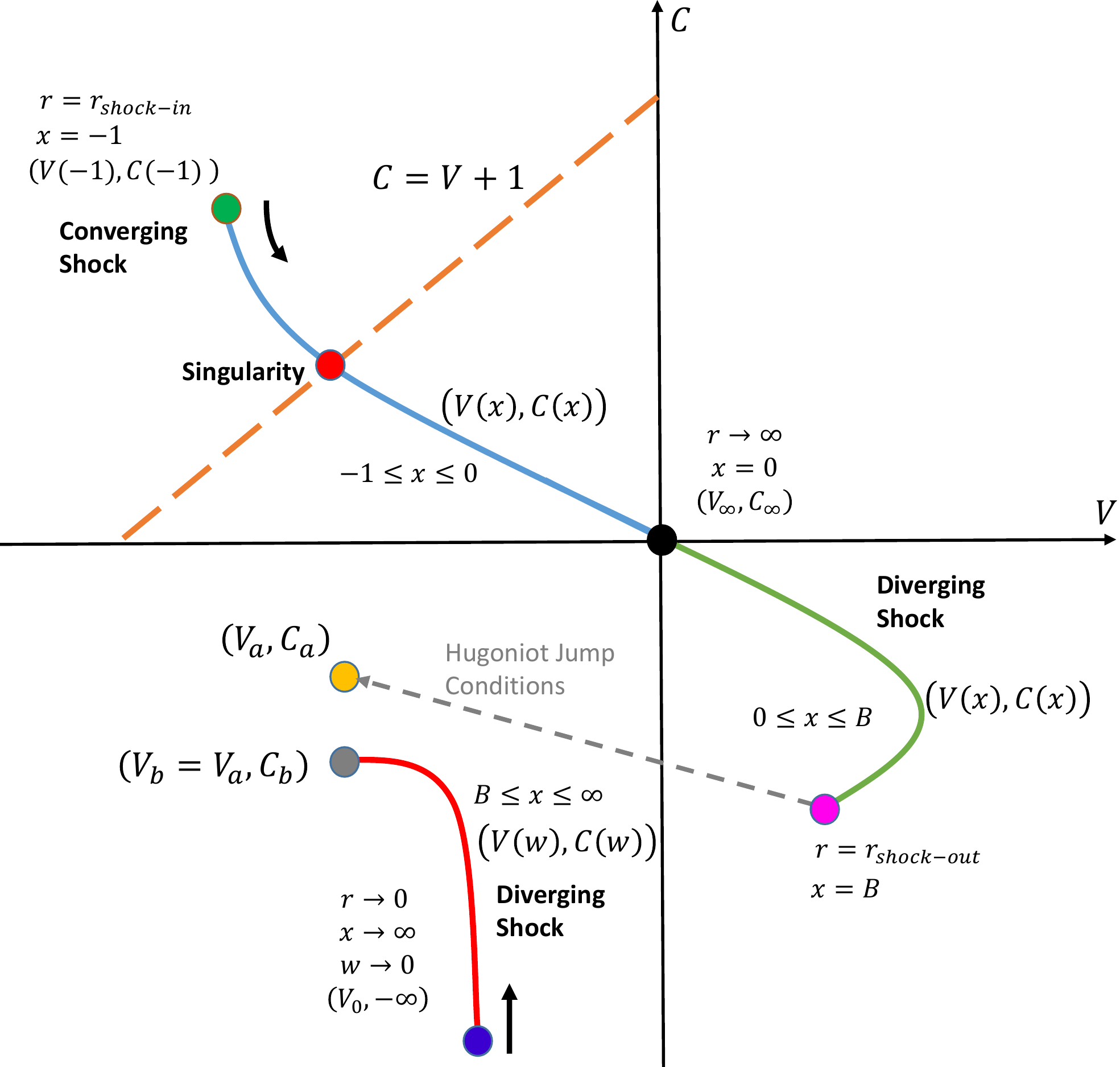}
\par\end{centering}
\caption{The integration path of the similarity ODEs (equation \ref{eq:ODEs})
in the $C-V$ plane. The integration starts (green point) at the converging
shock front where $x=-1$ (equations \ref{eq:V-1}, \ref{eq:C-1}),
then goes through the singularity (which is removable by definition
of the similarity exponent $\lambda$, as detailed in Ref. \cite{giron2021solutions}),
and reaches $x=0$ ($r=\infty$ of the converging shock profile),
as described by the blue curve. At this point (black point, where
$x=V=C=0$), the integration switches to positive times ($x\protect\geq0$)
and represents the diverging shock. It extends (green curve) until
it reaches the diverging shock front at $x=B$ (magenta point). The
properties of material on the other side of this front are then found
through the Rankine-Hugoniot jump conditions (equations \ref{eq:hugo_V}-\ref{eq: hugo_C}),
and we obtain the point $\left(V_{a},C_{a}\right)$ (yellow point).
Finally, we define another integration variable $w=x^{-\sigma}$ (as
detailed in the text) and integrate the ODEs \ref{eq: dVdw}-\ref{eq: dCdw}
to this shock front, from\textbf{ }the origin $\left(V_{0},-\infty\right)$
(blue point, $V_{0}$ given by equation \ref{eq:V0_val}), and continue
(red curve) to the shock front until the value $V\left(w\right)=V_{a}$
is reached. This integration will end at some value $C\left(w\right)=C_{b}$,
(gray point), and the correct value of the parameter $B$ is such
that $C_{a}=C_{b}$. \label{fig:CV-plane-pptx}}
\end{figure}

In this section we describe a robust method for calculating the parameter
$B$ using iterative algorithms. We generalize the method of Lazarus
\cite{lazarus1977similarity,lazarus1981self}, which dealt with the
specific case of an initially homogeneous density ($\mu=0$), to a
general power law density (equation \ref{eq:rhopowlaw}). In section
\ref{subsec:The-Required-Integration} we describe the proper integration
path for equation \ref{eq:ODEs}, which in turn, gives a constraint
which enables the calculation of the constant $B$. This method resembles
the calculation of the similarity exponent $\lambda$, which is determined
by a constraint on the integration path for the incoming shock, as
it crosses a singular point \cite{giron2021solutions}.

\subsection{The Integration Path \label{subsec:The-Required-Integration}}

The full integration path for the converging and diverging shock in
the $C-V$ plane is outlined and described in detail in figure \ref{fig:CV-plane-pptx}.
As discussed above, the diverging shock we consider here was generated
by a previously converging Guderley shock. As a result, the full integration
path of equation \ref{eq:ODEs} should first pass through the integration
path of the incoming shock \cite{giron2021solutions}, and in particular
it should start from the incoming shock front coordinate, $x=-1$,
where the strong shock Rankine--Hugoniot relations give the following
values for the similarity profiles at the front \cite{giron2021solutions}:

\begin{equation}
V\left(-1\right)=-\frac{2}{\gamma+1},\label{eq:V-1}
\end{equation}
\begin{equation}
R\left(-1\right)=\frac{\gamma+1}{\gamma-1},
\end{equation}
\begin{equation}
C\left(-1\right)=\frac{\sqrt{2\gamma\left(\gamma-1\right)}}{\gamma+1}.\label{eq:C-1}
\end{equation}
The integration is continued all the way to $x=0$ where $\left(V,C\right)=\left(0,0\right)$.
At this point, the integration continues consistently to positive
$x$, which represents the diverging shock from $r\rightarrow\infty$
and inwards, and is evaluated at positive times. The integration terminates
at the diverging shock front, which (equations \ref{eq:xdef},\ref{eq: shock position})
corresponds to $x=B$. Therefore integration of equations \ref{eq:ODEs}
from the incoming shock front at $x=-1$ to $x=B$ will give the similarity
profiles through the diverging shock as well, terminating at where
the solution is discontinuous according to the Rankine--Hugoniot
jump relations (equations \ref{eq:massjump}-\ref{eq:energyjump}).
These relations can be written in terms of the similarity profiles,
by substituting equations \ref{eq:V_def}-\ref{eq:R_def} into jump
relations, as derived in detail in Appendix \ref{sec:Self-Similar-Hugonoit}
(see also Ref. \cite{lazarus1981self}). Again denoting and the indices
1 and 2 correspond to the pre-shocked and post-shocked components
t by $\left(R_{1},V_{1},C_{1}\right)$ and $\left(R_{2},V_{2},C_{2}\right)$,
respectively, the resulting Rankine-Hugoniot jump relations are:

\begin{equation}
1+V_{2}=\frac{\gamma-1}{\gamma+1}\left(1+V_{1}\right)+\frac{2C_{1}^{2}}{\left(\gamma+1\right)\left(1+V_{1}\right)},\label{eq:hugo_V}
\end{equation}
\begin{equation}
R_{2}\left(1+V_{2}\right)=R_{1}\left(1+V_{1}\right).\label{eq:hugo_R}
\end{equation}
\begin{equation}
C_{2}^{2}=C_{1}^{2}+\frac{\gamma-1}{2}\left(\left(1+V_{1}\right)^{2}-\left(1+V_{2}\right)^{2}\right),\label{eq: hugo_C}
\end{equation}
We note that that in the case of the diverging shock we need to solve
for nonzero pressure in the upstream material, since it has been affected
by the converging shock (which in itself has been treated as strong,
propagating into cold gas) (See figures \ref{fig:The-analytical-density-pressure}-\ref{fig:analytic_shock_pos}).
After the jump at $x=B$ via equations \ref{eq:hugo_V}-\ref{eq: hugo_C},
the integration path should continue from $x=B$ and $\left(R_{2},V_{2},C_{2}\right)$
to $x=\infty\ \left(r\rightarrow0,\ t>0\right)$.

In order to perform this integration numerically, we first calculate
the asymptotic values of $V,C$ for $x\rightarrow\infty$. From symmetry
arguments the velocity at the center must vanish, that is $u\left(r\rightarrow0,t\right)=0$.
As a result, from the definition of the velocity similarity profile
$V\left(x\right)$ (equation \ref{eq:V_def}) we find,

\begin{equation}
\lim_{x\rightarrow\infty}x^{-\frac{1}{\lambda}}V\left(x\right)=0.\label{eq:bound_1}
\end{equation}
Since the speed of sound is always finite and positive for $r\rightarrow0$,
it follows from the definition of $C\left(x\right)$ (equation \ref{eq:C_def})
that: 
\begin{equation}
\lim_{x\rightarrow\infty}C\left(x\right)=-\infty.\label{eq: bound 2}
\end{equation}
If we now write the ODE for $V\left(x\right)$ at $x\rightarrow\infty$,
we obtain: 
\[
\frac{dV}{dx}=\frac{C^{2}\left(nV+\frac{2\left(\lambda-1\right)-\mu}{\gamma}\right)-V\left(V+1\right)\left(V+\lambda\right)}{\lambda x\left(C^{2}-\left(V+1\right)^{2}\right)}
\]
\[
\approx\frac{C^{2}\left(nV+\frac{2\left(\lambda-1\right)-\mu}{\gamma}\right)}{\lambda xC^{2}}=\frac{nV+\frac{2\left(\lambda-1\right)-\mu}{\gamma}}{\lambda x}
\]
This results in a simple ODE whose solution is given by 
\begin{equation}
V\left(x\right)=V_{0}+Ax^{\frac{n}{\lambda}},\label{eq:Vxinf}
\end{equation}
where:

\begin{equation}
V_{0}=-\frac{2\left(\lambda-1\right)-\mu}{n\gamma},\label{eq:V0_val}
\end{equation}
and $A$ is an integration constant. From equation \ref{eq:bound_1}
and since $n\geq1$ it follows that we must have $A=0$, so we obtain
the result: 
\begin{equation}
\lim_{x\rightarrow\infty}V\left(x\right)=V_{0}.\label{eq:v0lim}
\end{equation}
Equations \ref{eq: bound 2},\ref{eq:v0lim} gives the asymptotic
behavior for $V,C$ for $x\rightarrow\infty$. Following Lazarus \cite{lazarus1977similarity,lazarus1981self},
in order to properly handle the integration near $x\rightarrow\infty$,
we perform the following change of variables: 
\begin{equation}
w=kx^{-\sigma},\label{eq:xwdef}
\end{equation}
with a specific value for $\sigma>0$ which will be determined shortly.
We use the relation $\lambda x\frac{d}{dx}=-\lambda\sigma w\frac{d}{dw}$
in equation \ref{eq:ODEs} and obtain ODEs for $C\left(w\right),V\left(w\right)$:

\begin{equation}
\lambda\sigma w\frac{dV}{dw}=-\frac{\Delta_{2}}{\Delta},\label{eq: dVdw}
\end{equation}
\begin{equation}
\lambda\sigma w\frac{dC}{dw}=-\frac{\Delta_{3}}{\Delta},\label{eq: dCdw}
\end{equation}
which depends on $\sigma$ but not on the value of $k$. According
to equation \ref{eq: bound 2}, we expand $C\left(w\right)$ around
$w=0$ (which corresponds to $x\rightarrow\infty$):

\begin{equation}
C\left(w\right)=...+\frac{C_{-2}}{w^{2}}+\frac{C_{-1}}{w}+C_{0}+C_{1}w...\label{eq: queue_C}
\end{equation}
Writing equation \ref{eq: dCdw} explicitly (using equations \ref{eq:delta},\ref{eq:delta3})
in the limit $w\rightarrow0$, leads to,

\[
\frac{dC}{dw}=-\frac{C\left(w\right)}{w\lambda\sigma}\left(1+\frac{2\left(\lambda-1\right)+\mu\left(\gamma-1\right)}{2\gamma\left(1+V\left(w\right)\right)}\right).
\]
Inserting the expansion \ref{eq: queue_C} and taking the limit $w\rightarrow0$
yields, 
\begin{align*}
...-\frac{2C_{-2}}{w^{3}}-\frac{C_{-1}}{w^{2}} & =-\frac{1}{\lambda\sigma}\left(1+\frac{2\left(\lambda-1\right)+\mu\left(\gamma-1\right)}{2\gamma\left(1+V_{0}\right)}\right)\\
 & \times\left(...+\frac{C_{-2}}{w^{3}}+\frac{C_{-1}}{w^{2}}+\frac{C_{0}}{w}\right).
\end{align*}
We see that at every order $l\leq-1$ we must have: 
\begin{equation}
lC_{l}=-\frac{1}{\lambda\sigma}\left(1+\frac{2\left(\lambda-1\right)+\mu\left(\gamma-1\right)}{2\gamma\left(1+V_{0}\right)}\right)C_{l}.\label{eq:ckkk}
\end{equation}
Hence only one value of $l$ can satisfy equation \ref{eq:ckkk} with
$C_{l}\neq0$, depending on the value of $\sigma$. As a result, by
taking: 
\begin{equation}
\sigma=\frac{1}{\lambda}\left(1+\frac{2\left(\lambda-1\right)+\mu\left(\gamma-1\right)}{2\gamma\left(1+V_{0}\right)}\right),\label{eq:sigdef}
\end{equation}
we set $l=-1$ with $C_{-1}\neq0$, while $C_{l}=0$ for all $l\leq-2$.
Finally, the value of $k$ can thus be set such that $C_{-1}=1$.
In summary, with the choice of equation \ref{eq:sigdef}, we have:

\begin{equation}
C\left(w\rightarrow0\right)\approx-\frac{1}{w}\label{eq: queue_C-1}
\end{equation}

\subsection{Numerical bracketing of $B$ \label{subsec:The-Algorithm}}

\begin{figure}
\begin{centering}
\includegraphics[scale=0.42]{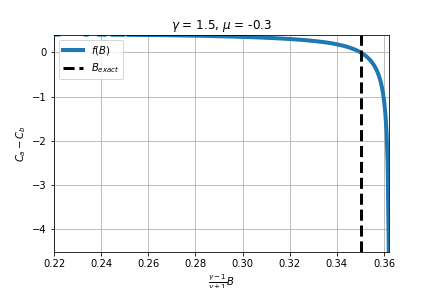}
\par\end{centering}
\caption{The difference between the dimensionless sound speed from both sides
of the integration (blue curve), as a function of $\frac{\gamma-1}{\gamma+1}B$
(as described in Sec. \ref{subsec:The-Algorithm}), for $\gamma=1.5$,
$\mu=-0.3$, $n=3$. This difference vanishes for the correct value
of $B$ (vertical black line). \label{fig:iterate_B}}
\end{figure}

The key to the full solution of the diverging shock is to determine
the correct value of the parameter $B$, which we do iteratively.
As described in figure \ref{fig:CV-plane-pptx} which depicts the
integration path in the $C-V$ plane, for a given trial value of $B$
the ODEs \ref{eq:ODEs} are integrated from $x=-1$ to $x=B$, and
then using the jump conditions we obtain $V_{2},C_{2}$ from equations
\ref{eq:hugo_V},\ref{eq: hugo_C} which we denote as $\left(V_{a},C_{a}\right)$.

In the next step, the ODEs are integrated from the other direction
($x\rightarrow\infty$) towards $x=B$. This is done by switching
the integration variable to $w$ (equation \ref{eq:xwdef}) and integrating
equations \ref{eq: dVdw}-\ref{eq: dCdw} from some point $w_{0}\ll1$
with the initial values $\left(V,C\right)=\left(V_{0},\frac{1}{w_{0}}\right)$,
according to the asymptotic relations \ref{eq:Vxinf},\ref{eq: queue_C-1}.
In practice we take $w_{0}=10^{-7}$. The integration is carried out
until the value $V\left(w\right)=V_{a}$ is reached. The value of
$C\left(V_{a}\right)$ obtained at the end of the integration on $w$
is denoted by $C_{b}$. If the correct value of $B$ is used, the
integration from both sides will meet at the same point in the $C-V$
plane (figure \ref{fig:CV-plane-pptx}), that is, we must have $C_{a}=C_{b}$.

This method allows for a robust calculation of $B$, by numerically
bracketing the root of the function $f\left(B\right)=C_{a}-C_{b}$,
as shown in figure \ref{fig:iterate_B}. In order to set an initial
range $\left[B_{\text{min}},B_{\text{max}}\right]$ for the the bracketing,
we found that 
for a wide range of $\gamma$ and $\mu$: 
\begin{equation}
0\leq\frac{\gamma-1}{\gamma+1}B\left(\gamma,\mu,n\right)\leq2
\end{equation}
(this formula is an extension of that found by Lazarus \cite{lazarus1981self}
for $\mu=0$, with $1$ on the right hand side of the inequality).

In order to set the initial bracketing range $\left[B_{\text{min}},B_{\text{max}}\right]$
such that $f\left(B_{\text{min}}\right)\cdot f\left(B_{\text{max}}\right)<0$,
we linearly scan the segment $\left[0,2\frac{\gamma+1}{\gamma-1}\right]$
until we find $B_{\text{min}},B_{\text{max}}$ which result in different
signs of $f\left(B\right)$. After finding $B_{\text{max}}$ and $B_{\text{min}}$
we use a standard bisection method to bracket the root $B$. Some
caution is necessary, since $C_{a}-C_{b}$ may diverge for certain
values of $B$, as shown in figure \ref{fig:iterate_B}. If such divergence
occurs, we reduce the scanning increment $\Delta B$ and repeat the
process from the last successful iteration, $B_{*}$, until once again
we find an appropriate range of $\left[B_{\text{min}},B_{\text{max}}\right]$.

\subsection{Integration of the Self-Similar Profiles \label{subsec:The-Self-Similar-Profiles}}

\begin{figure*}[t]
\begin{centering}
\includegraphics[scale=0.47]{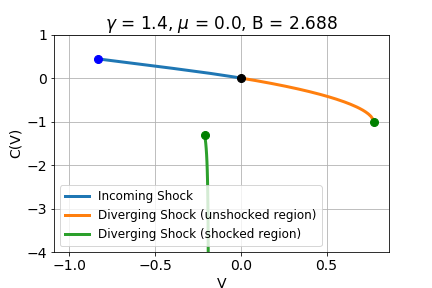}\includegraphics[scale=0.47]{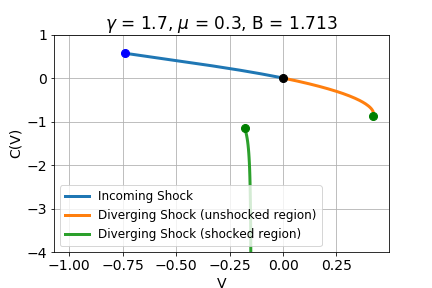}
\par\end{centering}
\begin{centering}
\includegraphics[scale=0.47]{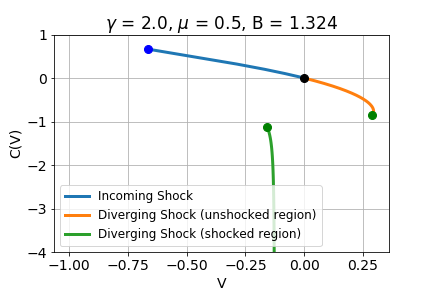}\includegraphics[scale=0.47]{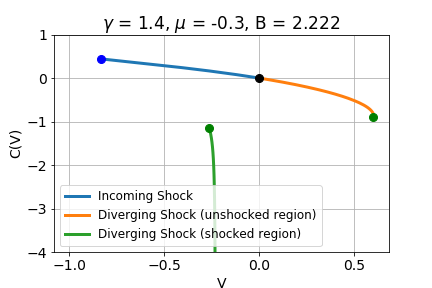}
\par\end{centering}
\caption{The curves $C\left(V\right)$ of four different cases in spherical
symmetry - the parameters and the resulting values of $B$ are listed
in the titles. The integration of $C\left(V\right)$ in the $C-V$
plane (as described in detail in figure \ref{fig:CV-plane-pptx}),
is divided into 3 regions: (i) the blue curve represents the converging
shock from the shock front to $r\rightarrow\infty$, (ii) the orange
curve represents the upstream flow from $r\rightarrow\infty$ to the
diverging shock front and (iii) the green curve represents the downstream
flow from the diverging shock front to the center $r=0$. \label{fig:CV_cases}}
\end{figure*}

\begin{figure}
\begin{centering}
\includegraphics[scale=0.5]{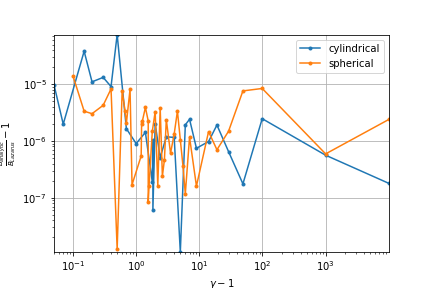}
\par\end{centering}
\caption{The relative error between the values of $B$ calculated with our
algorithm and values found by Lazarus for $\mu=0$ \cite{lazarus1981self},
as a function of $\gamma-1$, for cylindrical (blue) and spherical
(orange) symmetries. \label{fig:comp Lazarus}}
\end{figure}

\begin{figure}
\begin{centering}
\includegraphics[scale=0.5]{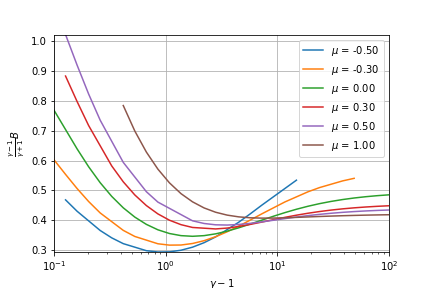}
\par\end{centering}
\caption{$\frac{\gamma-1}{\gamma+1}B$ as a function of $\gamma$ for different
values of $\mu$ (as listed in the legend) and in spherical symmetry.
\label{fig:-as-a}}
\end{figure}

\begin{table}
\begin{centering}
\begin{tabular}{|c|c|c|}
\hline 
\multicolumn{3}{|c|}{$\gamma=\frac{5}{3}$}\tabularnewline
\hline 
$\mu$ & $B_{n=2}$ & $B_{n=3}$\tabularnewline
\hline 
$-1.0$ & $0.83233629$ & $0.92183037$\tabularnewline
\hline 
$-0.684210$ & $1.03899820$ & $1.08282717$\tabularnewline
\hline 
$-0.526315$ & $1.16539327$ & $1.17631586$\tabularnewline
\hline 
$0.263157$ & $2.02484184$ & $1.76890317$\tabularnewline
\hline 
$0.736842$ & $2.74231348$ & $2.23013019$\tabularnewline
\hline 
\end{tabular}
\par\end{centering}
\begin{centering}
\begin{tabular}{|c|c|c|}
\hline 
\multicolumn{3}{|c|}{$\gamma=1.4$}\tabularnewline
\hline 
$\mu$ & $B_{n=2}$ & $B_{n=3}$\tabularnewline
\hline 
$-1.0$ & $1.14084671$ & $1.38075188$\tabularnewline
\hline 
$-0.684210$ & $1.54284404$ & $1.71998084$\tabularnewline
\hline 
$-0.526315$ & $1.78569523$ & $1.91403860$\tabularnewline
\hline 
$0.263157$ & $3.47396741$ & $3.15677630$\tabularnewline
\hline 
$0.4210526$ & $3.92133697$ & $3.46678822$\tabularnewline
\hline 
\end{tabular}
\par\end{centering}
\caption{Various values of $B$ for $\gamma=\frac{5}{3},1.4$ for different
values of $\mu$ for cylindrical ($n=2$) and spherical ($n=3$) symmetries.\label{tab:Canonical}}
\end{table}

\begin{figure*}[t]
\begin{centering}
\includegraphics[scale=0.47]{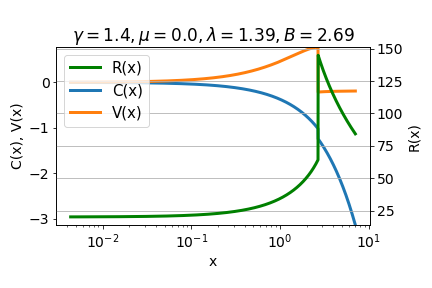}\includegraphics[scale=0.47]{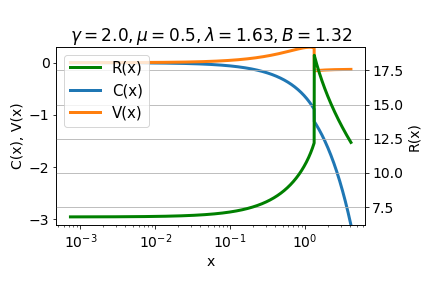}
\par\end{centering}
\begin{centering}
\includegraphics[scale=0.47]{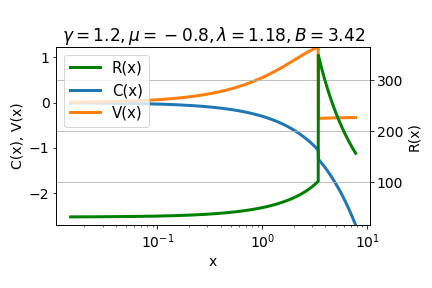}\includegraphics[scale=0.47]{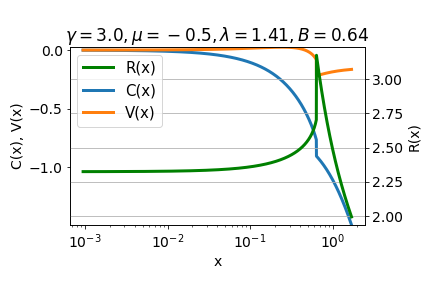}
\par\end{centering}
\caption{The self-similar profiles $V\left(x\right)$ (orange, left $y$-axis),
$C\left(x\right)$ (blue, left $y$-axis) and $R\left(x\right)$ (green,
right $y$-axis) as a function of $x$, for four different cases in
spherical symmetry - the parameters and the resulting values of $B,\lambda$
are listed in the titles. \label{fig:The-self-similar-profiles}}
\end{figure*}

As described in figure \ref{fig:CV-plane-pptx}, after the correct
value of $B$ is determined, we can integrate \ref{eq:ODEs} to obtain
the semi-analytic self-similar profiles $V\left(x\right),C\left(x\right),R\left(x\right)$.
The integration starts from $x=-1$ with the (converging) strong shock
conditions: 
\[
\left(V,C,R\right)\Bigg|_{x=-1}=\left(-\frac{2}{\gamma+1},\frac{\gamma+1}{\gamma-1},\frac{\sqrt{2\gamma\left(\gamma-1\right)}}{\gamma+1}\right),
\]
and continues to $x=0$ to the values representing the diverging shock,
$t>0$, and $x\geq0$. From this point the solution represents the
diverging shock from $r\rightarrow\infty$ to the diverging shock
front. At the shock front $x=B$, we use the jump conditions \ref{eq:hugo_R}-\ref{eq: hugo_C}
and continue the integration to $x\rightarrow\infty$ $\left(r\rightarrow0\right)$.
In practice we integrate on a finite range $x\leq x_{\infty}$, which
is truncated at: 
\[
x_{\infty}=\frac{t}{r_{\text{min}}^{\lambda}}
\]
where we take $r_{\text{min}}=0.05$.

\subsection{Results \label{subsec:Results}}

In figure \ref{fig:CV_cases} we plot the curve $C\left(V\right)$
for several cases, as obtained from the numerical integration described
in figure \ref{fig:CV-plane-pptx}. To the best of our knowledge,
numerical results for the values of $B$ appear in the literature
only for $\mu=0$. In this case, the values published by Lazarus \cite{lazarus1981self}
are used as a benchmark to evaluate the algorithm - as was done for
example in Ref. \cite{ramsey2012guderley}. Thus we compare our results
for the calculation of $B\left(\gamma,\mu=0,n\right)$ in figure \ref{fig:comp Lazarus},
for a wide range of $\gamma$ and for spherical ($n=3$) and cylindrical
symmetries ($n=2$) . Except for a few cases, the agreement is better
than $10^{-5}$, an mostly to all digits given Ref \cite{lazarus1981self}.
In figure \ref{fig:-as-a} we present the value $\frac{\gamma-1}{\gamma+1}B$
as a function of $\gamma$ for various values of $\mu$. In table
\ref{tab:Canonical} we list the numerical values of $B$ for $\gamma=1.4,\frac{5}{3}$
for several values of $\mu$. Finally, In figure \ref{fig:The-self-similar-profiles}
the self similar profiles $V\left(x\right),C\left(x\right),R\left(x\right)$
are shown for several examples.

\section{Comparison with Hydrodynamic simulations \label{sec:Comparison-to-a}}

\begin{figure*}[t]
\begin{centering}
\includegraphics[scale=0.47]{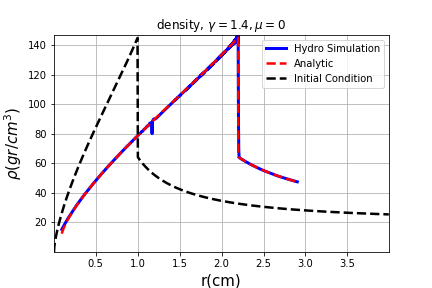}\includegraphics[scale=0.47]{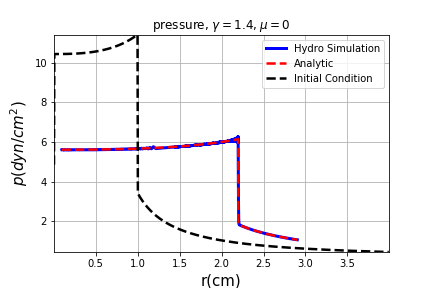}
\par\end{centering}
\begin{centering}
\includegraphics[scale=0.47]{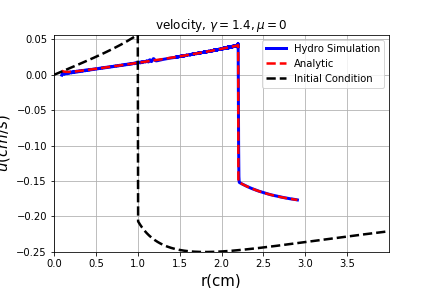}\includegraphics[scale=0.47]{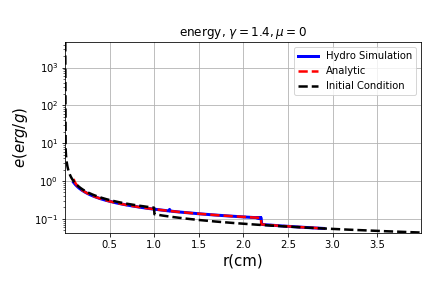}
\par\end{centering}
\caption{A comparison between the analytical solution (dashed red lines) and
a numerical hydrodynamic simulation (blue lines) for $\gamma=1.4$,
$\mu=0$, $n=3$ of the hydrodynamic profiles (as listed in the titles),
for the initially diverging shock setup, in which the simulation is
initialized with the analytic profiles shown in the dashed black lines
at time $t=B=2.688$, and evolves to time $t=2B$. The simulation
was carried out with $1000$ computational cells.\label{fig:simulation_out}}
\end{figure*}

\begin{figure*}[t]
\begin{centering}
\includegraphics[scale=0.47]{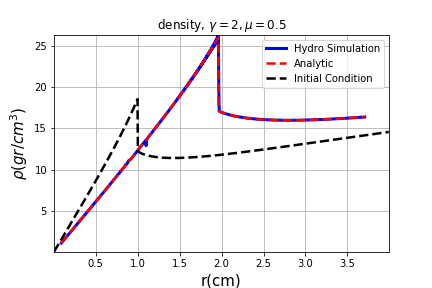}\includegraphics[scale=0.47]{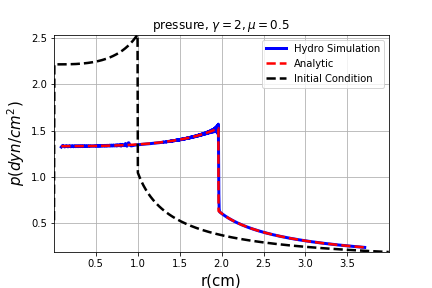}
\par\end{centering}
\begin{centering}
\includegraphics[scale=0.47]{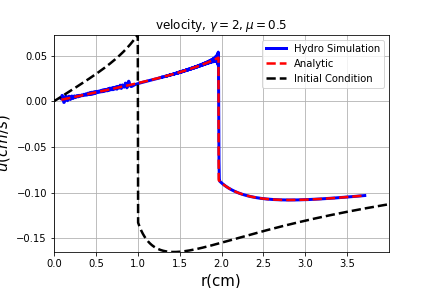}\includegraphics[scale=0.47]{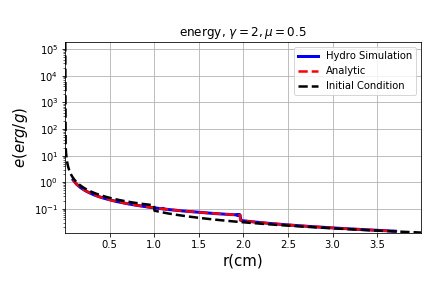}
\par\end{centering}
\caption{Same as figure \ref{fig:simulation_out}, for the case $\gamma=2,\mu=0.5$,
for which the initial time is $t=B=1.324$.\label{fig:simulation_out-1}}
\end{figure*}

\begin{figure*}[t]
\begin{centering}
\includegraphics[scale=0.47]{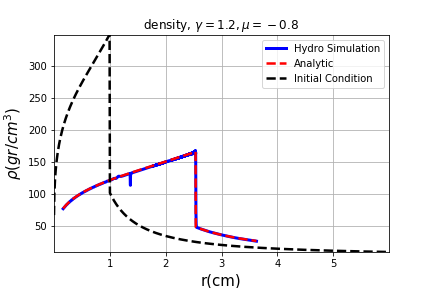}\includegraphics[scale=0.47]{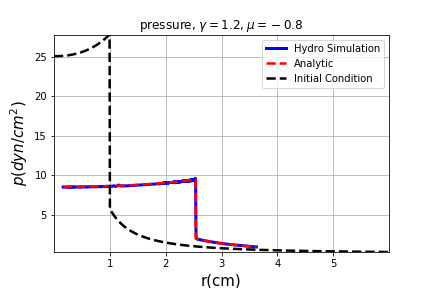}
\par\end{centering}
\begin{centering}
\includegraphics[scale=0.47]{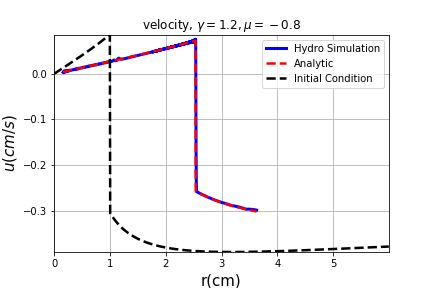}\includegraphics[scale=0.47]{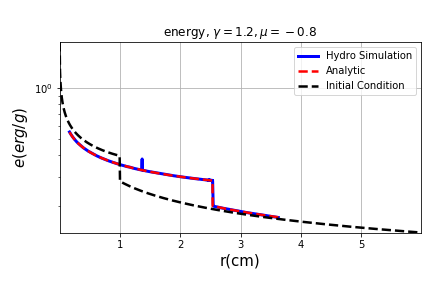}
\par\end{centering}
\caption{Same as figure \ref{fig:simulation_out}, for the case $\gamma=1.2,\mu=-0.8$,
for which the initial time is $t=B=3.418$.\label{fig:simulation_out-2}}
\end{figure*}

\begin{figure*}[t]
\begin{centering}
\includegraphics[scale=0.42]{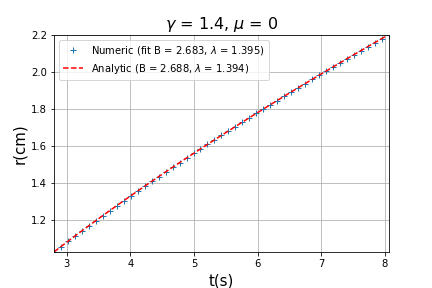}\includegraphics[scale=0.42]{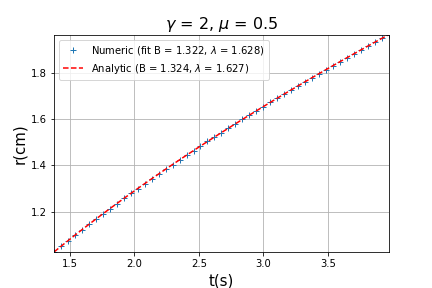}\includegraphics[scale=0.42]{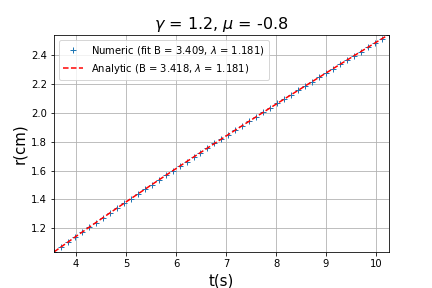}
\par\end{centering}
\caption{A comparison for the shock position as a function of time, between
the analytical solution (red dashed line) and hydrodynamic simulations
(crosses), for the simulations presented in figures \ref{fig:simulation_out}-\ref{fig:simulation_out-2}
(as listed in the titles). We list in the legends the analytic and
simulated results for the values of $B,\lambda$, which are obtained
from the simulation by fitting to shock position to $\left(t/B\right)^{\frac{1}{\lambda}}$.
\label{fig:shockpos_out}}
\end{figure*}

\begin{figure*}[t]
\begin{centering}
\includegraphics[scale=0.42]{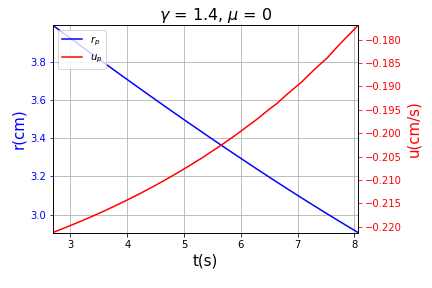}\includegraphics[scale=0.42]{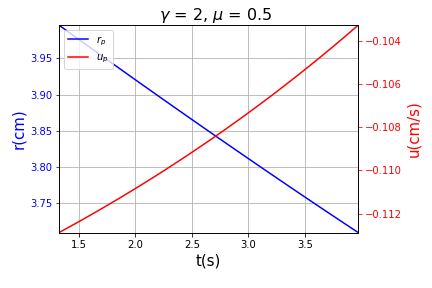}\includegraphics[scale=0.42]{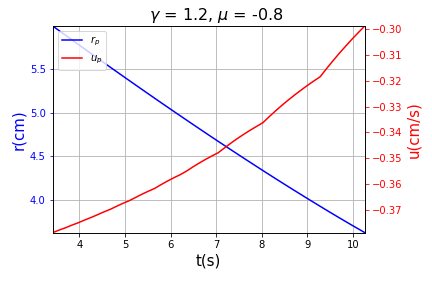}
\par\end{centering}
\caption{The piston boundary position (blue line, left axis) and applied velocity
(red line, right axis) as a function of time, for the simulations
presented in figures \ref{fig:simulation_out}-\ref{fig:simulation_out-2}
(as listed in the titles).\label{fig:bc_out}}
\end{figure*}

\begin{figure*}[t]
\begin{centering}
\includegraphics[scale=0.42]{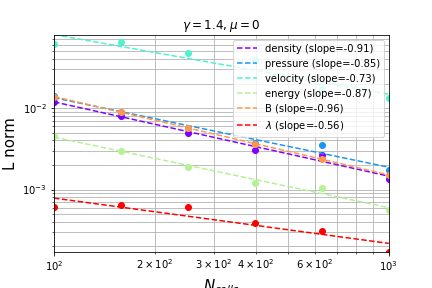}\includegraphics[scale=0.42]{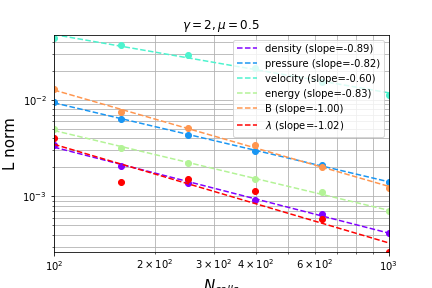}\includegraphics[scale=0.42]{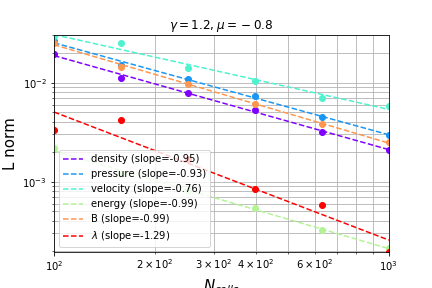}
\par\end{centering}
\caption{Numerical convergence of various quantities as a function of the number
of cells for the cases presented in figures \ref{fig:simulation_out}-\ref{fig:simulation_out-2}
(as listed in the titles). Shown are the relative $L_{1}$ norms (see
equation \ref{eq:lnorm}) for the density (purple), pressure (blue),
velocity (cyan) and energy (green) profiles, as well as the relative
errors for the fitted parameters $B$ (orange) and $\lambda$ (red).
The points represent actual data and the dashed lines represent a
log-log linear fit, whose resulting slope is listed in the legends.
This slope represents the order of spatial convergence.\label{fig:lnorm_out}}
\end{figure*}

\begin{figure*}[t]
\begin{centering}
\includegraphics[scale=0.21]{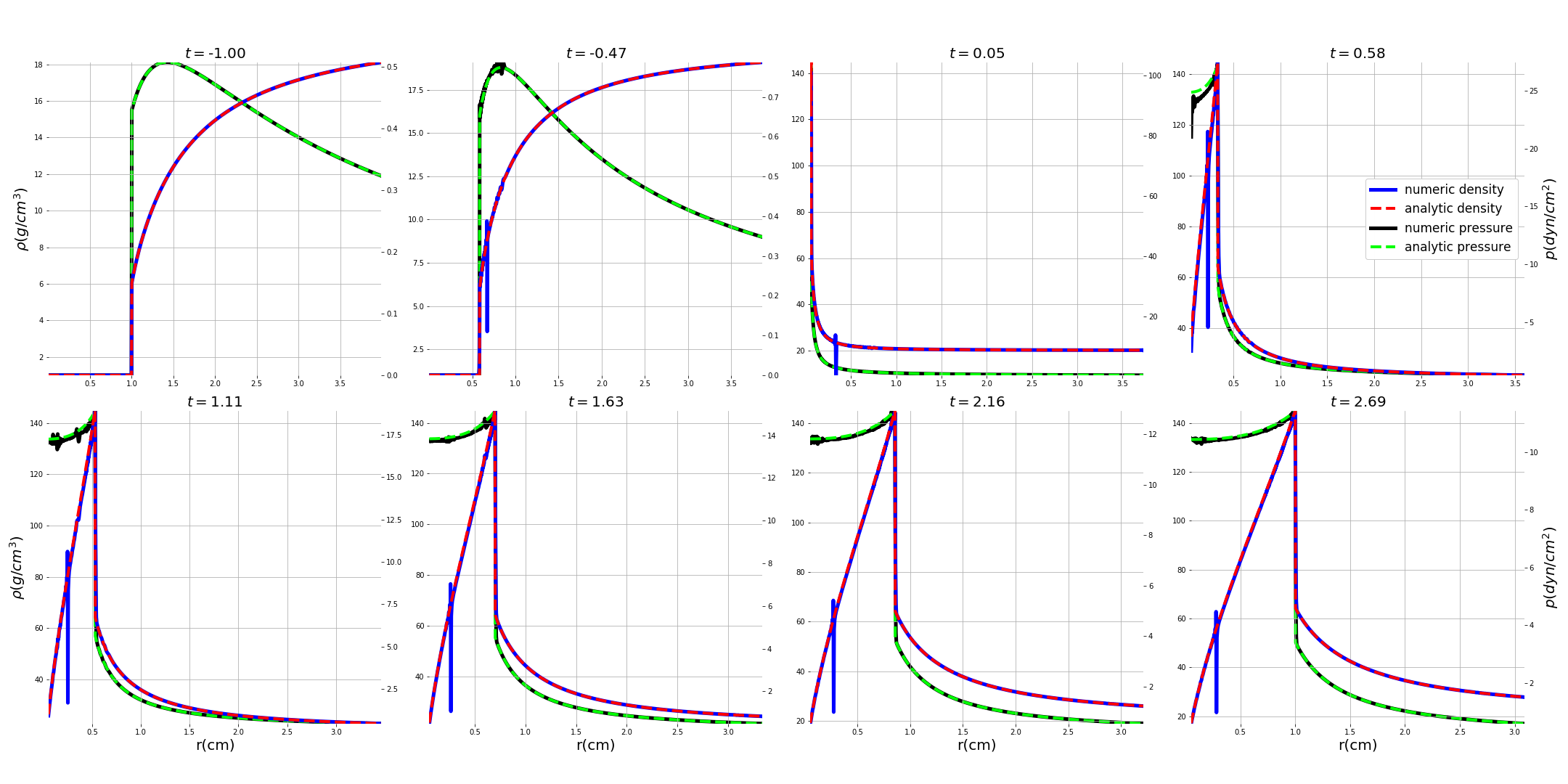}
\par\end{centering}
\caption{A comparison between the analytical solution and a numerical hydrodynamic
simulation (with 1000 cells) for the density (left $y$ axis, analytic
- dashed red line, simulation - solid blue line) and pressure (right
y axis, analytic - dashed green line, simulation - solid black line)
profiles for the converging shock setup and for $\gamma=1.4$, $\mu=0$.
The simulation is initialized with the analytic converging shock profiles
at time $t=-1$ where the converging shock front is at $r=1$. We
let the flow evolve as the shock reaches the center at time $t=0$
and reflects outwards, until it reaches the position $r=1$ at the
final time $t=B=2.684$ (see also figure \ref{fig:shock_pos_inout}).
The simulation was carried with $1000$ computational cells. \label{fig:simulation_inout_1}}
\end{figure*}

\begin{figure*}[t]
\begin{centering}
\includegraphics[scale=0.21]{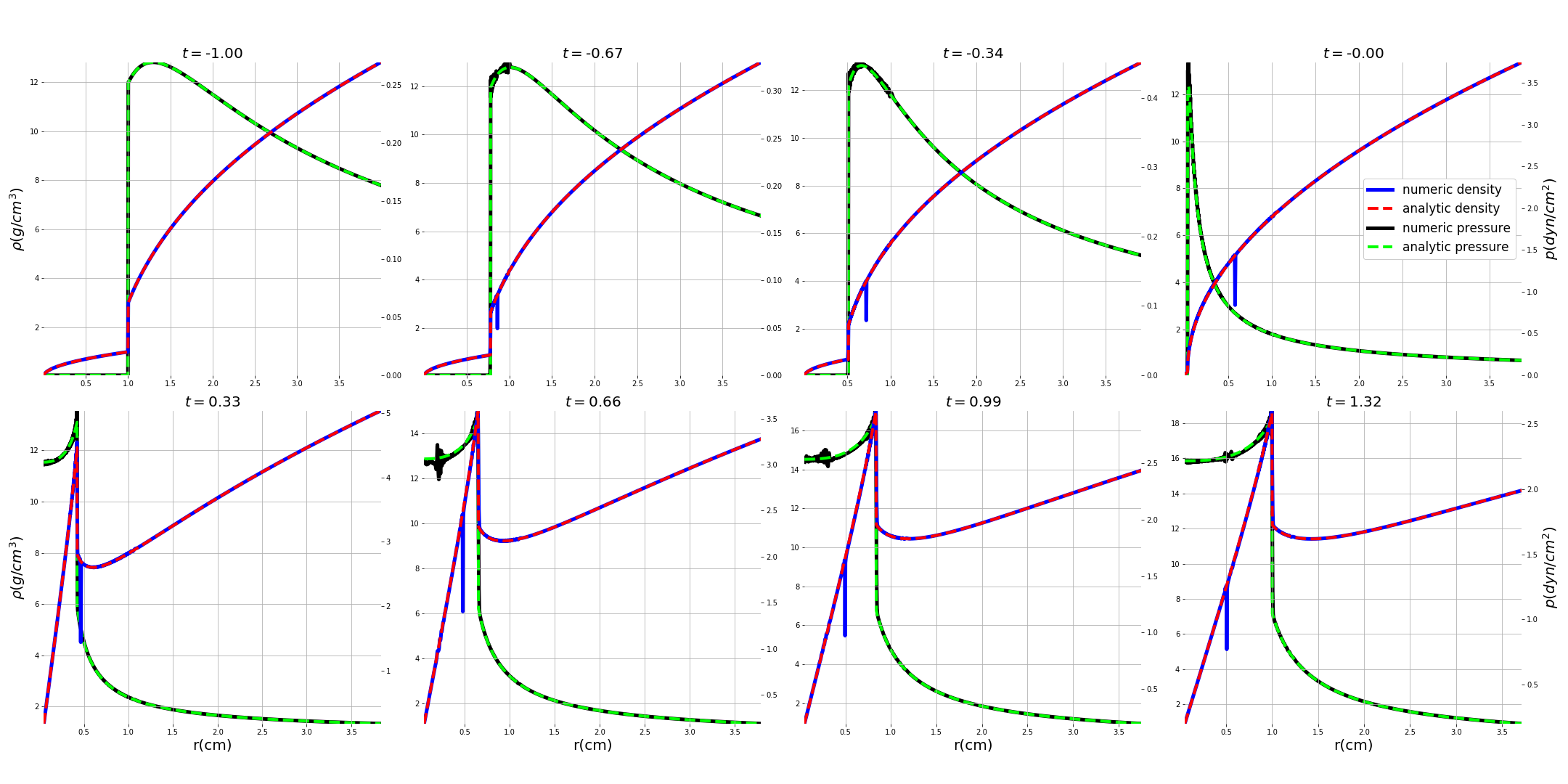}
\par\end{centering}
\caption{Same as figure \ref{fig:simulation_inout_1}, for the case $\gamma=2,\mu=0.5$,
for which the final time is $t=B=1.324$.\label{fig:simulation_inout_2}}
\end{figure*}

\begin{figure*}[t]
\begin{centering}
\includegraphics[scale=0.21]{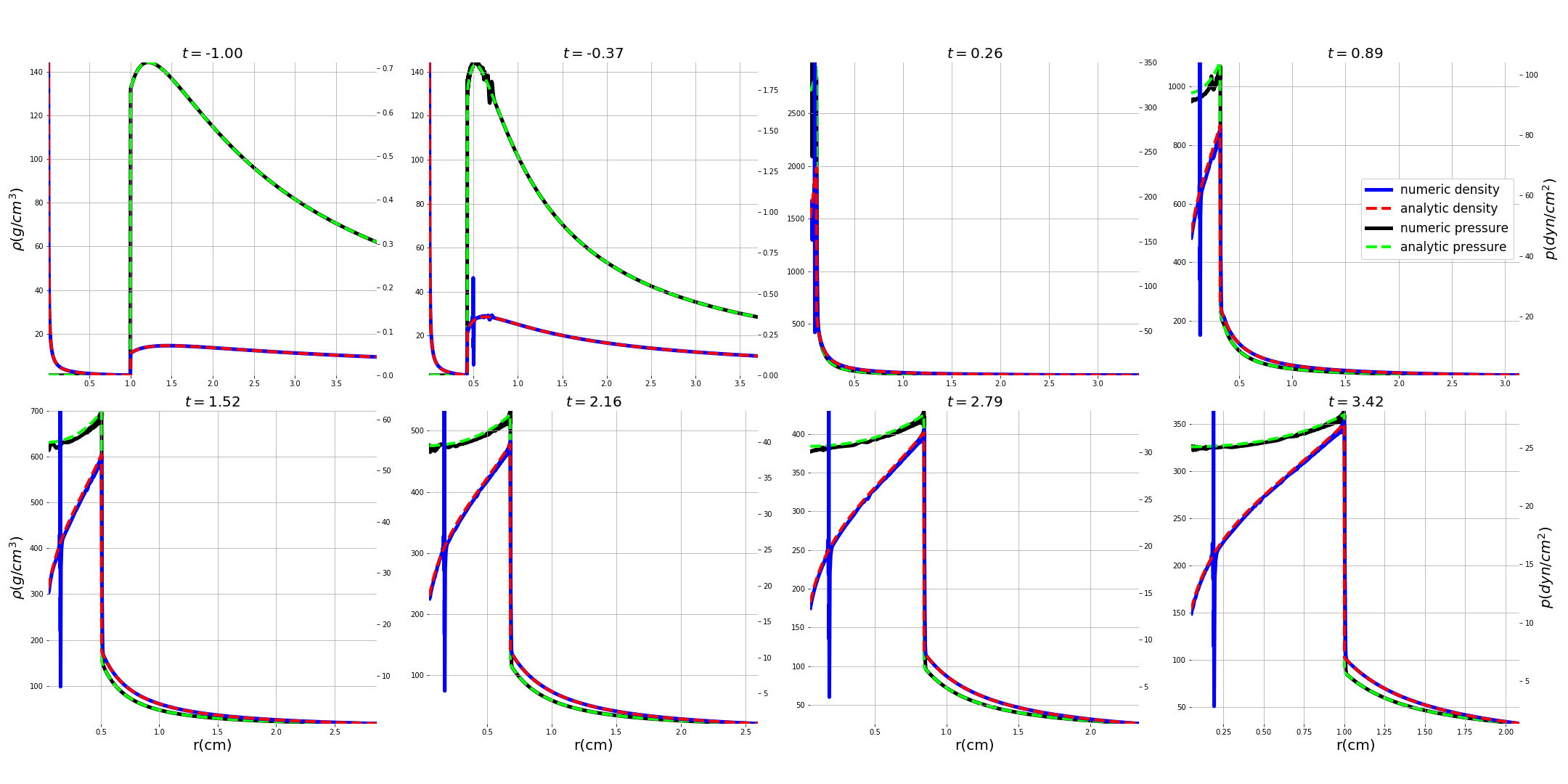}
\par\end{centering}
\caption{Same as figure \ref{fig:simulation_out}, for the case $\gamma=1.2,\mu=-0.8$,
for which the final time is $t=B=3.418$ .\label{fig:simulation_inout_3}}
\end{figure*}

\begin{figure*}[t]
\begin{centering}
\includegraphics[scale=0.42]{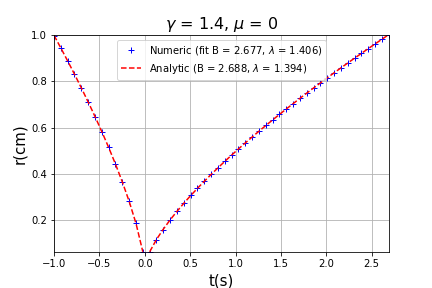}\includegraphics[scale=0.42]{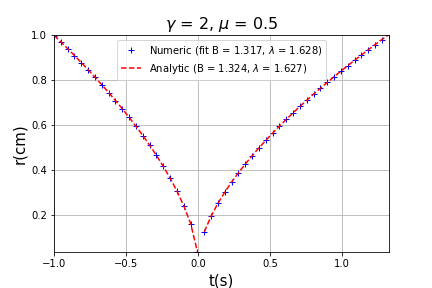}\includegraphics[scale=0.42]{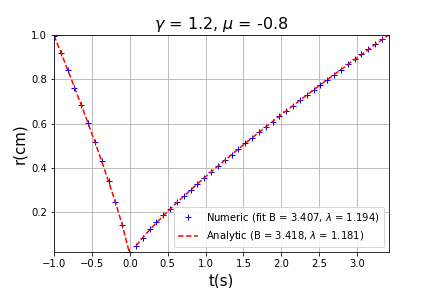}
\par\end{centering}
\caption{A comparison for the shock position as a function of time, between
the analytical solution (red dashed line) and hydrodynamic simulations
(crosses), for the simulations presented in figures \ref{fig:simulation_inout_1}-\ref{fig:simulation_inout_3}
(as listed in the titles). We list in the legends the analytic and
simulated results for the values of $B,\lambda$, which are obtained
from the simulation by fitting to diverging shock position ($t>0$)
to $\left(t/B\right)^{\frac{1}{\lambda}}$.\label{fig:shock_pos_inout}}
\end{figure*}

\begin{figure*}[t]
\begin{centering}
\includegraphics[scale=0.42]{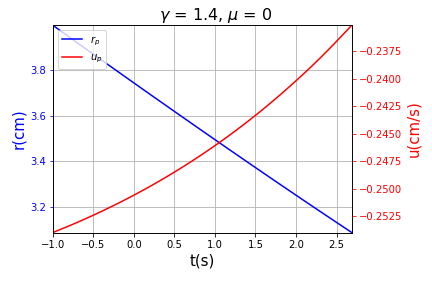}\includegraphics[scale=0.42]{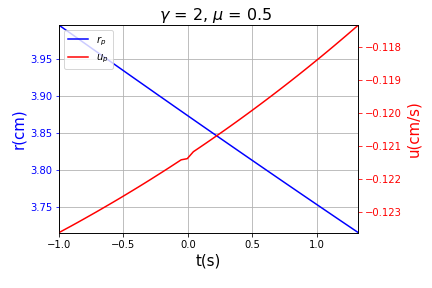}\includegraphics[scale=0.42]{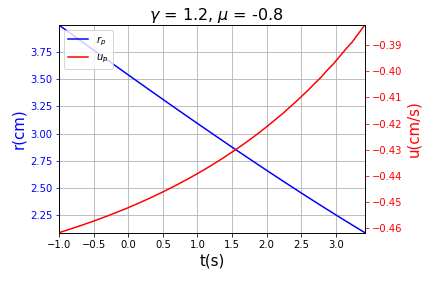}
\par\end{centering}
\caption{The piston boundary position (blue line, left axis) and applied velocity
(red line, right axis) as a function of time, for the simulations
presented in figures \ref{fig:simulation_inout_1}-\ref{fig:simulation_inout_3}
(as listed in the titles).\label{fig:bc_inout}}
\end{figure*}

\begin{figure*}[t]
\begin{centering}
\includegraphics[scale=0.42]{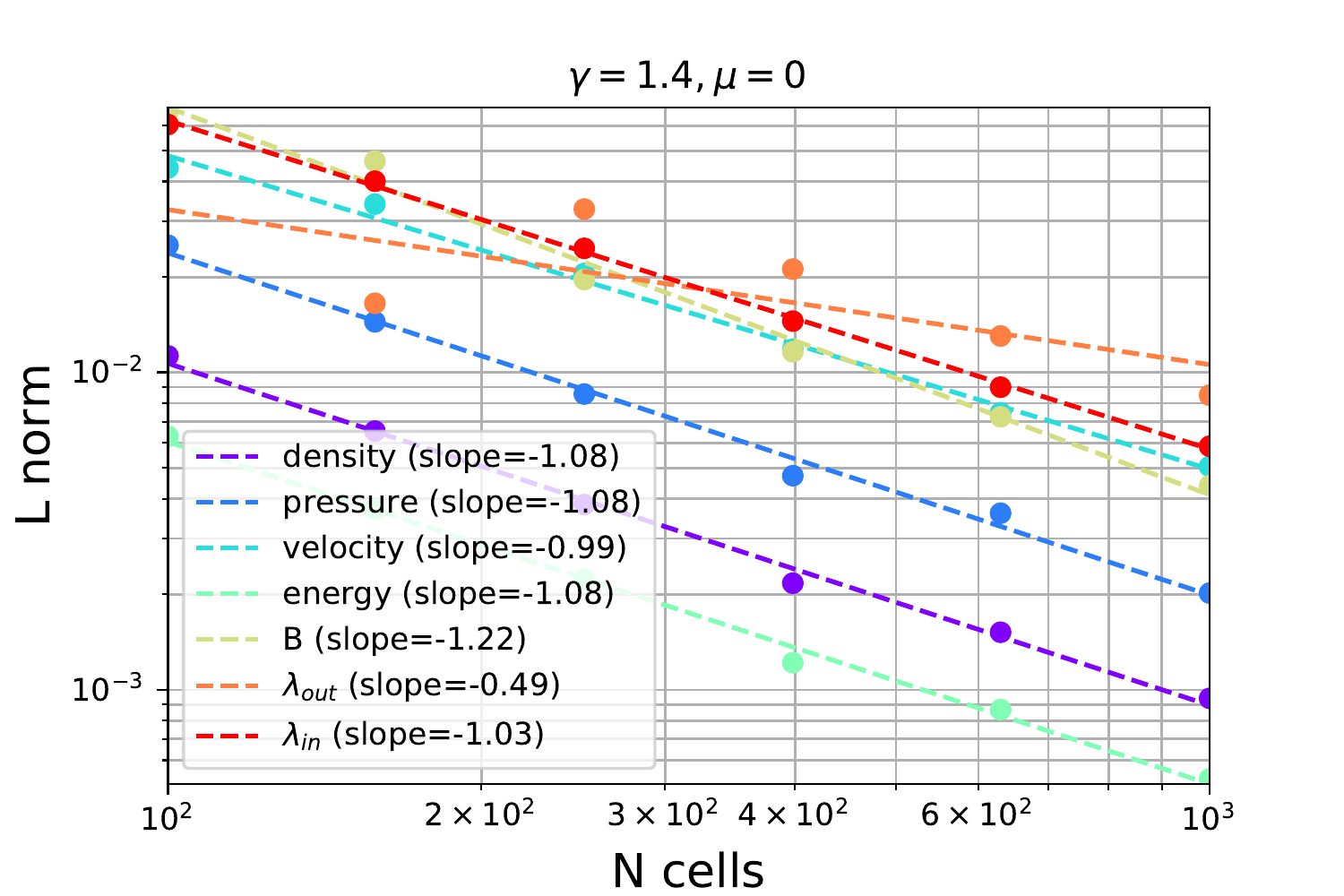}\includegraphics[scale=0.42]{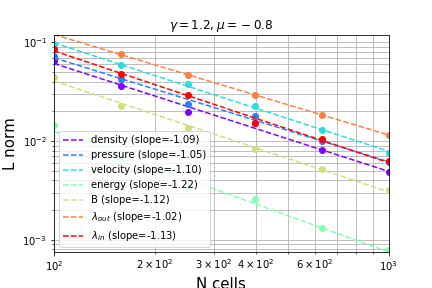}\includegraphics[scale=0.42]{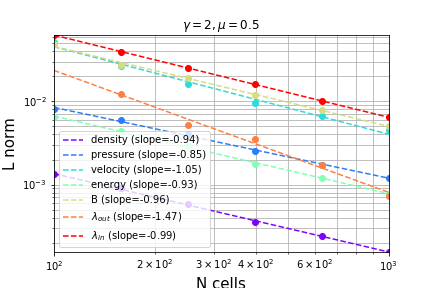}
\par\end{centering}
\caption{Numerical convergence of various quantities as obtained from the simulation
to the analytical solution, a function of the number of cells for
the cases presented in figures \ref{fig:simulation_inout_1}-\ref{fig:simulation_inout_3}
(as listed in the titles). Shown are the relative$L_{1}$ norms (see
equation \ref{eq:lnorm}) for the density (purple), pressure (blue),
velocity (cyan) and energy (green) profiles, as well as the relative
errors for the fitted diverging shock position parameters $B$ (yellow)
and $\lambda_{\text{out}}$ (orange) and the converging shock parameter
$\lambda_{\text{in}}$(red), according to equation \ref{eq: shock position}.
The points represent actual data and the dashed lines represent a
log-log linear fit, whose resulting slope is listed in the legends.
This slope represents the order of spatial convergence. \label{fig:lnorm_inout}}
\end{figure*}

In this section we compare the semi-analytical diverging shock solutions
to numerical hydrodynamic simulations. As in Ref. \cite{giron2021solutions},
we employ a standard staggered-mesh artificial-viscosity one-dimensional
Lagrangian code (for details, see Appendix D. in Ref. \cite{giron2021solutions}). 

We define two types of setups: (i) an initialization of a diverging
shock flow, which propagates outwards and (ii) an initialization of
an incoming shock flow which first converges to the origin and reflects
outwards as a diverging shock which propagates outward into the converging
shock flow. The latter case tests the ability of the simulation to
accurately describe both the focusing of the shock at the center and
the generation of the reflected shock, which then propagates into
the incoming shock profile. The former case is more simple, as it
does not include the shock focusing and reflection, but is still non-trivial
as it describes the propagation of the outgoing shock.

As was done in the case for the converging shock \cite{giron2021solutions},
we employ a piston velocity boundary condition, where the velocity
is obtained from the analytic solution for the (converging and diverging)
Guderley problem \cite{ramsey2019piston}, that is, if the location
of the piston is $r_{p}\left(t\right)$, its velocity which is externally
applied is given by: 
\[
u_{p}\left(t\right)=u_{\text{analytic}}\left(r_{p}\left(t\right),t\right),
\]
where $u_{\text{analytic}}$ is the analytic velocity profile (equation
\ref{eq:V_def}) and the position of the piston is calculated in the
Lagrangian simulation as the position of the outermost cell vertex.
All simulations are initialized on a spatially uniform radial grid.

The hydrodynamics profiles are compared using relative $L_{p}$ norms
which are defined as follows: 
\begin{equation}
L_{p}\left(y,y^{\text{analytic}},v\right)=\frac{\left(\sum_{i}v_{i}\left|y_{i}-y_{i}^{\text{analytic}}\right|^{p}\right)^{\frac{1}{p}}}{\left(\sum_{i}v_{i}\left|y_{i}\right|^{p}\right)^{\frac{1}{p}}+\left(\sum_{i}v_{i}\left|y_{i}^{\text{analytic}}\right|^{p}\right)^{\frac{1}{p}}},\label{eq:lnorm}
\end{equation}
where $v_{i}$ is the volume of cell $i$, and $y_{i}$, $y_{i}^{\text{analytic}}$
represent, respectively, the numerical and (semi)analytical hydrodynamic
profiles.

We define 3 test cases: 
\begin{enumerate}
\item $\gamma=1.4,\ \mu=0$ - this is the canonical Guderley problem with
a uniform initial density profile. 
\item $\gamma=2,\ \mu=0.5$ - a test case with a monotonically outward-increasing
density profile. While the converging shock accelerates inward in
such a profile, We expect the diverging shock to decelerate, as it
advances into a denser material. 
\item $\gamma=1.2,\ \mu=-0.8$ - a test case with a monotonically decreasing
density profile. In this case the opposite occurs: the converging
shock decelerates on its way inwards, and the diverging shock accelerates,
as it advances into an increasingly rarefied material. 
\end{enumerate}
Notably, a sharp discontinuity appears in the hydrodynamic profiles
in all three cases (figures \ref{fig:simulation_out}-\ref{fig:simulation_out-2}
and \ref{fig:simulation_inout_1}-\ref{fig:simulation_inout_3}).
As discussed in Ref. \cite{giron2021solutions}, these discontinuities
exist over a single computational cell and were originated due to
the presence of discontinuities in the initial hydrodynamic profiles
\cite{leveque2002finite}. As seen in the figures, they do not affect
the overall stability and accuracy of the simulations. A detailed
analysis of such initialization errors can be found in Ref.\textbf{
}\cite{ramsey2012guderley}. 

As in Ref. \cite{giron2021solutions}, we see that while there exists
some variability in the accuracy of the simulations regarding the
different physical quantities, we observe that a grid of 1000 computational
cells will generally suffice to ensure an accuracy of a few percent
or even less than one percent in all physical quantities at the end
of the simulation. The spatial convergence rates (reduction of error
w.r.t. the analytical solution as a function of increase in grid size)
is similar for all physical quantities in all simulations. These rates
are all of order unity (as seen explicitly in figures \ref{fig:lnorm_out}
and \ref{fig:lnorm_inout}), which is to be expected for an artificial
viscosity numerical scheme, which is first order accurate in the presence
of shocks.

\subsection{Initialization as a Diverging Shock}

As in the converging shock setup defined in Refs. \cite{ramsey2012simulation,ramsey2017verification,giron2021solutions},
we initialize the simulation with density, velocity and pressure profiles
given by the analytical diverging solution at time $t=B$, where the
diverging shock is at $r=1$. We let the flow evolve until the time
$t=2B$.

For each case we present the following plots: hydrodynamic profiles
at the final time $t=2B$ (figures \ref{fig:simulation_out}-\ref{fig:simulation_out-2}),
the shock position (figure \ref{fig:shockpos_out}), the applied piston
boundary condition (figure \ref{fig:bc_out}), the convergence of
$L_{1}$ norms for $\rho,p,u,e$ (figure \ref{fig:lnorm_out}), and
the convergence of $\lambda,B$ which are calculated from the simulation
using a fit of the shock position to the form $\left(\frac{t}{B}\right)^{\frac{1}{\lambda}}$. 

\subsection{Initialization as a Converging Shock}

In this setup, we initialize the simulation with density, velocity
and pressure profiles given by the analytical converging solution
at time $t=-1$, where the converging shock is at $r=1$. We let the
flow evolve as the shock reaches the center at time $t=0$ and reflects
outwards, until it reaches the position $r=1$ at time $t=B$. For
each case we present the following plots: density and pressure profiles
at the final time $t=B$ (figures \ref{fig:simulation_inout_1}-\ref{fig:simulation_inout_3}),
the converging and diverging shock position as function of time (figure
\ref{fig:shock_pos_inout}), the applied piston boundary condition
(figure \ref{fig:bc_inout}), the convergence of $L_{1}$ norms for
$\rho,p,u,e$ (figure \ref{fig:lnorm_inout}), and the convergence
of the diverging shock parameters $\lambda_{\text{out}},B$ which
are calculated from the simulation using a fit of the diverging shock
position to the form $\left(\frac{t}{B}\right)^{\frac{1}{\lambda}}$
for $t>0$, and the converging shock exponent $\lambda_{\text{in}}$,
which is calculated from the simulation using a fit of the converging
shock position to the form $\left(-t\right)^{\frac{1}{\lambda}}$
for $t<0$. 

\section{Summary \label{sec:Summary}}

In this work we studied in detail the complete Guderley problem of
a converging and then diverging shock in an ideal gas medium with
an initial power-law density profile, $\rho\left(r\right)\sim r^{\mu}$.
We extend our previous work in Ref. \cite{giron2021solutions}, which
was dedicated to a detailed study of the converging shock. We presented
the theoretical framework required to construct exact self-similar
solutions for the converging and diverging flows. A generalized algorithm
for a numerical calculation of the semi-analytical similarity solutions
was developed. The diverging shock solution was studied in a wide
range of parameters, namely the adiabatic constant $\gamma$, the
initial density exponent $\mu$ and for spherical and cylindrical
symmetries. The results are in excellent agreement with previous published
work, which, to the best of our knowledge, are only given for the
case of an initially constant density profile ($\mu=0$).

The semi-analytic solutions were used to define nontrivial compressible
flow problems which could serve for code verification. The solutions
were compared to hydrodynamic simulations employing a one dimensional
Lagrangian hydrodynamic code, by applying appropriate initial and
boundary conditions. A very good agreement was reached between numerical
hydrodynamic simulations and the semi-analytical solutions. This success
highlights the use of the solutions for the purpose of verification
and validation of numerical hydrodynamics simulation codes.

\subsection*{Availability of data}

The data that support the findings of this study are available from
the corresponding author upon reasonable request.

\bibliographystyle{unsrt}
\bibliography{datab}

\appendix

\section{Rankine-Hugoniot Jump Relations \label{sec:Self-Similar-Hugonoit}}

In this appendix we use the similarity representation \ref{eq:V_def}-\ref{eq:R_def},
in order to write the dimensional Rankine-Hugoniot jump relations
\ref{eq:massjump}-\ref{eq:energyjump} for the diverging shock, in
terms of the similarity profiles. We note that the location of the
diverging shock varies according to the power law $r_{s}\left(t\right)=\left(\frac{t}{B}\right)^{\frac{1}{\lambda}}$.
Hence the shock velocity in the jump relations \ref{eq:massjump}-\ref{eq:energyjump}
is $D=\frac{dr_{s}}{dt}=\frac{r_{s}}{\lambda t}$.

The mass equation \ref{eq:massjump} reads: 
\[
\frac{\rho_{0}r_{s}^{1+\mu}}{\lambda t}\left(R_{2}-R_{1}\right)=\frac{\rho_{0}r_{s}^{1+\mu}}{\lambda t}R_{1}V_{1}-\frac{\rho_{0}r_{s}^{1+\mu}}{\lambda t}R_{2}V_{2},
\]
where, as expected, all dimensional quantities drop, and we obtain
the first dimensionless jump condition:

\begin{equation}
R_{2}\left(1+V_{2}\right)=R_{1}\left(1+V_{1}\right).
\end{equation}

Dividing the momentum equation \ref{eq:momentumjump} by the continuity
equation \ref{eq:massjump} and writing the pressure in terms of the
speed of sound, we get after some straightforward algebra: 
\[
u_{1}-u_{2}=\frac{1}{\gamma}\left(\frac{c_{2}^{2}}{u_{2}-D}-\frac{c_{1}^{2}}{u_{1}-D}\right).
\]
Next, expressing this equation in terms of the specific enthalpy $h=e+\frac{p}{\rho}+\frac{1}{2}\left(u-D\right)^{2}$,
which is constant across the shock, we obtain: 
\begin{equation}
\frac{1}{\left(u_{1}-D\right)\left(u_{2}-D\right)}=\frac{\gamma+1}{2\left(\gamma-1\right)h}.
\end{equation}
This last relation is then used for $u_{2}-D$:
\[
u_{2}-D=\frac{\gamma-1}{\gamma+1}\left(u_{1}-D\right)+\frac{2c_{1}^{2}}{\left(\gamma+1\right)\left(u_{1}-D\right)}.
\]
Finally, we substitute the shock velocity and use the similarity representation
\ref{eq:V_def}-\ref{eq:C_def}:

\[
-\frac{r_{s}}{\lambda t}V_{2}-\frac{r_{s}}{\lambda t}=\frac{\gamma-1}{\gamma+1}\left(-\frac{r_{s}}{\lambda t}V_{1}-\frac{r_{s}}{\lambda t}\right)+\frac{2\left(-\frac{r_{s}}{\lambda t}C_{1}\right)^{2}}{\left(\gamma+1\right)\left(-\frac{r_{s}}{\lambda t}V_{1}-\frac{r_{s}}{\lambda t}\right)}
\]
As expected, all dimensional quantities drop and we obtain the second
dimensionless jump condition: 
\begin{equation}
1+V_{2}=\frac{\gamma-1}{\gamma+1}\left(1+V_{1}\right)+\frac{2C_{1}^{2}}{\left(\gamma+1\right)\left(1+V_{1}\right)}.
\end{equation}

In order to obtain the third dimensionless jump condition, we write:
\begin{equation}
\frac{p}{\rho}=\frac{c^{2}}{\gamma}=\frac{r^{2}C^{2}}{\gamma\lambda^{2}t^{2}},
\end{equation}
and derive for the specific energy 
\begin{equation}
e=\frac{p}{\left(\gamma-1\right)\rho}=\frac{r^{2}C^{2}}{\gamma\left(\gamma-1\right)\lambda^{2}t^{2}}.
\end{equation}
Substituting this result in the energy jump condition (equation \ref{eq:energyjump})
leads to: 
\begin{align*}
 & \frac{r_{s}^{2}C_{1}^{2}}{\gamma\left(\gamma-1\right)\lambda^{2}t^{2}}+\frac{r_{s}^{2}C_{1}^{2}}{\gamma\lambda^{2}t^{2}}+\frac{1}{2}\left(-\frac{r_{s}}{\lambda t}V_{1}-\frac{r_{s}}{\lambda t}\right)^{2}\\
 & =\frac{r_{s}^{2}C_{2}^{2}}{\gamma\left(\gamma-1\right)\lambda^{2}t^{2}}+\frac{r_{s}^{2}C_{2}^{2}}{\gamma\lambda^{2}t^{2}}+\frac{1}{2}\left(-\frac{r_{s}}{\lambda t}V_{2}-\frac{r_{s}}{\lambda t}\right)^{2}
\end{align*}
and as before, all dimensional quantities drop and we obtain the third
dimensionless jump condition: 
\begin{equation}
C_{2}^{2}=C_{1}^{2}+\frac{\gamma-1}{2}\left(\left(1+V_{2}\right)^{2}-\left(1+V_{1}\right)^{2}\right).
\end{equation}

\end{document}